\documentclass[a4paper,onecolumn,11pt,accepted=2017-05-09]{quantumarticle}
\pdfoutput=1
\usepackage[utf8]{inputenc}
\usepackage[english]{babel}
\usepackage[T1]{fontenc}
\usepackage{amsmath,amssymb}
\usepackage{hyperref}
\usepackage[numbers,sort&compress]{natbib}
\usepackage{tikz,lipsum,orcidlink,bm}

\newcommand{\dd}{\mathrm{d}}
\newcommand{\ee}{\mathrm{e}}
\newcommand{\Tr}{\mathop{\mathrm{Tr}} \nolimits}
\newcommand{\R}{\mathop{\mathrm{Re}}\nolimits}
\newcommand{\I}{\mathop{\mathrm{Im}}\nolimits}
\newcommand{\op}[1]{\hat{#1}}
\begin{document}
	
	\title{Wigner negativity and stellar rank for SU(1,1) states}
	
	\author{Andrei~B.~Klimov\orcidlink{0000-0001-8493-721X}}
	\affiliation{Departamento de F\'{\i}sica, Universidad de Guadalajara, 44420~Guadalajara,  Jalisco, Mexico}
	
	\author{Ariana~Mu\~{n}oz\orcidlink{0000-0002-5971-7968}}
	\affiliation{Facultad de Ingenier\'{\i}a, Universidad Autónoma de Chile, Talca 3460000, Chile}
	\affiliation{Departamento de \'Optica, Facultad de F\'{\i}sica, Universidad Complutense, 28040~Madrid, Spain}
	
	\author{Gerd~Leuchs\orcidlink{0000-0003-1967-2766}}
	\affiliation{Max-Planck-Institut f\"{u}r die Physik des Lichts, 91058~Erlangen, Germany}  
	\affiliation{Friedrich-Alexander-Universit\"{a}t Erlangen-N\"{u}rnberg, 91058~Erlangen, Germany}
	
	\author{Jean-Pierre Gazeau\orcidlink{0000-0001-7681-7672}}
	\affiliation{Universit\'e Paris Cit\'e, CNRS, Astroparticule et Cosmologie, 75013 Paris, France}
	\affiliation{Faculty of Mathematics, University of Bia\l ystok, 15-245 Bia\l ystok, Poland}
	
	\author{Luis~L.~S\'{a}nchez-Soto\orcidlink{0000-0002-7441-8632}}
	\affiliation{Departamento de \'Optica, Facultad de F\'{\i}sica, Universidad Complutense, 28040~Madrid, Spain}
	\affiliation{Max-Planck-Institut f\"{u}r die Physik des Lichts, 91058~Erlangen, Germany}  
	\affiliation{Institute for Quantum Studies, Chapman University, Orange, CA 92866, USA}

	\maketitle
	
	\begin{abstract}
		Quasiprobability distributions for systems endowed with SU(1,1) dynamical symmetry have received surprisingly little attention, despite the central role of this symmetry in  two-photon physics, squeezed states, and nonlinear interferometry. Here, we fill this gap by constructing a  full covariant family of  $s$-ordered  quasiprobability distributions defined  on the two-sheeted hyperboloid, or equivalently, on the Poincaré unit disk via stereographic projection. A key result is that the Wigner function is strictly positive for all Perelomov SU(1,1) coherent states, in sharp contrast to the SU(2) case. This positivity endows Wigner negativity with an unambiguous operational meaning: any negative volume is a direct  signature of genuinely quantum behavior.  We further examine the stellar rank of SU(1,1) states, defined through the zeros of the Husimi $Q$-function, and show how it compares with Wigner negativity as a geometry-adapted witness of nonclassicality in this setting. We further introduce a hierarchy of generalized multipoles through a harmonic expansion of the density operator on the hyperboloid, providing a complementary framework for probing quantumness. This offers a comprehensive toolkit for characterizing and quantifying quantum resources in SU(1,1) systems.
	\end{abstract}
	
\section{Introduction}

The quantumness of a state is fundamentally connected to the symmetry structure of the underlying system and, in particular, to the set of admissible physical operations. The phase-space formalism~\cite{Schroek:1996fv,Schleich:2001hc,QMPS:2005mi,Weinbub:2018aa} offers a natural framework for exploring this connection. A consistent construction is possible whenever the symmetry group acts irreducibly on the Hilbert space, which guarantees a well-defined correspondence between quantum states and distributions in an appropriate homogeneous phase space~\cite{Kirillov:2004aa}.

Coherent states~\cite{Perelomov:1986kl,Gazeau:2009aa,Robert:2021aa} play a pivotal role in this setting. Generated by the action of the symmetry group on a fiducial state, they serve as the natural  classical  reference states of the theory. Labeling them with points in phase space yields a systematic covariant correspondence between operators in Hilbert space and functions on the phase space, recasting quantum mechanics in a classical language and providing a natural arena for studying the quantum-to-classical transition.

Within this framework, quantumness is identified with deviations from the behavior associated with the convex set generated by coherent states. This aligns naturally with modern quantum resource theories~\cite{Streltsov:2017aa,Chitambar:2019aa}, where coherent states constitute the free, symmetry-preserving states and quantumness measures quantify the operational advantage (metrological, computational, or cryptographic) that a state can provide over them.

For continuous-variable systems, governed by Heisenberg-Weyl symmetry, several phase-space criteria have been developed to characterize quantumness~\cite{Dodonov:2002aa}. Early examples include the Mandel $Q$-parameter~\cite{Mandel:1979aa}, which captures deviations of photon-number statistics from Poissonian behavior, and the Glauber-Sudarshan $P$ representation~\cite{Glauber:1963aa,Sudarshan:1963aa}, whose failure to exist as a regular probability density signals genuine quantum features~\cite{Titulaer:1965aa,Tan:2020aa}. More recent approaches emphasize geometric structure, such as the stellar rank~\cite{Chabaud:2020aa}, defined by the zeros of the Husimi $Q$-function~\cite{Husimi:1940aa} and equivalent to the Majorana constellation~ \cite{Majorana:1932aa}. Among these criteria, the negativity of the Wigner function~\cite{Kenfack:2004lw} remains the most widely used; for pure states, it coincides with non-Gaussianity according to Hudson's theorem~\cite{Hudson:1974kb}. Such Wigner-negative states have been realized experimentally in optical cavities~\cite{Yoshikawa:2013aa}, trapped ions~\cite{Fluhmann:2019aa}, and superconducting circuits~\cite{Lu:2021aa}, establishing them as a practical resource for quantum information processing.

This picture, however, is not universal. For systems associated with other symmetry groups, Wigner negativity loses its privileged status as a classicality criterion. The special role of Gaussian states on the real line does not carry over to more general phase-space geometries, and coherent states need not yield positive Wigner functions. In SU(2) systems, whose phase space is the Bloch sphere~\cite{Stratonovich:1956aa,Berezin:1975aa,Varilly:1989bh}, even coherent states can exhibit Wigner negativity, so that positivity acquires meaning only in the large-spin limit~\cite{Klimov:2017aa,Davis:2021aa}. An analogous complication arises for the Euclidean group E(2), with the cylinder as phase space~\cite{Mukunda:1979uq,Plebanski:2000fk,Rigas:2011by,Kastrup:2016aa,Fabre:2023aa}.
Thus, the interpretation of the Wigner negativity must be adapted to the symmetry of the systems, and geometry of the related phase space.

The situation changes fundamentally for SU(1,1). Here, as we show, the hyperbolic geometry of the phase space restores a clean separation: coherent states always yield a strictly positive Wigner function, so that any Wigner negativity is an unambiguous sufficient witness of nonclassicality relative to the convex hull of coherent states. This is a qualitatively stronger statement than what is available for SU(2) or E(2), and it gives Wigner negativity a well-defined operational meaning for the full family of SU(1,1) systems.

This distinction is particularly significant given the central role of SU(1,1) symmetry in two-photon physics~\cite{Wodkiewicz:1985aa,Gerry:1985aa,Gerry:1991aa,Gerry:1995kq,Gazeau:2023aa,Olmo:2020aa} and the recent experimental realization of nonlinear SU(1,1) interferometers~\cite{Jing:2011aa,Hudelist:2014aa}, first proposed by Yurke \textit{et al.}~\cite{Yurke:1986yg} , whose quantum-enhanced phase sensitivity makes the characterization of their nonclassical resources practically urgent~\cite{Chekhova:2016aa}. Yet, despite this relevance, phase-space treatments of SU(1,1) have remained sparse. Early work by Orlowski and Wodkiewicz~\cite{Orowski:1990aa} addressed specific quasidistributions for squeezed states, while Alonso \textit{et al.}~\cite{Alonso:2002aa} constructed Wigner functions on hyperboloids within a broader mathematical framework. What has
been lacking is a systematic analysis of what phase-space negativity means for quantumness in this setting. The present work fills this gap.

Specifically, we analyze the complete family of $s$-parametrized quasiprobability distributions on the two-sheeted hyperboloid, or equivalently on the Poincar\'{e}\ unit disk via stereographic projection, and prove that the Wigner function is strictly positive for all SU(1,1) coherent states. We further use a complementary perspective based on harmonic functions on the hyperboloid. In the SU(2) setting, such functions reduce to spherical harmonics and their expansion coefficients define the state multipoles~\cite{Hoz:2013aa}; quantities that encode the full information of a quantum state while remaining directly observable~\cite{Goldberg:2022aa}. Here we propose the SU(1,1) analogue of this multipole hierarchy and use it to contrast states whose information is concentrated in low-order multipoles with those spread across higher-order components~\cite{Goldberg:2020aa,Goldberg:2024aa}. Together, Wigner negativity and multipole structure provide two complementary and geometry-adapted lenses through which the quantumness of SU(1,1) states can be characterized and quantified.

\section{Basic settings}

\label{sec:basic}

We consider a quantum system endowed with SU(1,1) dynamical symmetry, meaning that the Hamiltonian of the system can be expressed exclusively in terms of the Lie algebra $\mathfrak{su}(1,1)$. This algebra is spanned by the operators $\{\hat{K}_{0},\hat{K}_{1},\hat{K}_{2}\}$ with commutation relations (in units $\hbar =1$ throughout) 
\begin{equation}
[ \hat{K}_{1},\hat{K}_{2} ]=-i \hat{K}_{0}\,, \qquad  [ \hat{K}_{2}, \hat{K}_{0}]=i\hat{K}_{1}\,,
\qquad  [\hat{K}_{0},\hat{K}_{1}]=i\hat{K}_{2} \, .  
\label{eq:CR1}
\end{equation}
It is often convenient to introduce the ladder operators $\hat{K}_{\pm }= \hat{K}_{1}\pm i\hat{K}_{2}$, in terms of which~\eqref{eq:CR1} becomes 
\begin{equation}
[ \hat{K}_{0}, \hat{K}_{\pm }]=\pm \hat{K}_{\pm } \,,\qquad 
 [\hat{K}_{+},\hat{K}_{-}]=-2\hat{K}_{0}\,  \
 \label{eq:CR2}
\end{equation}
These relations closely resemble those of $\mathfrak{su}(2)$, differing only by a crucial sign. This sign change underlies the distinct physical properties of SU(1,1) systems compared with their SU(2) counterparts.

The Casimir operator, the analog of the total angular momentum, reads now 
\begin{equation}
\hat{K}^{2}=\hat{K}_{0}^{2}-\hat{K}_{1}^{2}-\hat{K}_{2}^{2}=\hat{K}_{0}^{2}-
\frac{1}{2}(\hat{K}_{+}\hat{K}_{-}+\hat{K}_{-}\hat{K}_{+})=k(k-1)\openone,
\end{equation}
where the eigenvalue $k$, known as the Bargmann index, labels the irreducible representation (irreps), which are infinite dimensional. There are several different series of irreps for SU(1,1) fixed by the domains of the eigenvalues $k$~\cite{Bargmann:1947fk}. We will focus on the positive discrete series, where $k=\tfrac{1}{2},1,\tfrac{3}{2},2,\ldots  $  (the index $k=1/2$ is treated as a limiting case).

The carrier space of the irrep labeled by $k$ is the Hilbert space $\mathcal{D}_{k}^{+}$ spanned by the orthonormal basis $\{|k,\mu \rangle \,| \mu =k,k+1,k+2,\ldots \}$ consisting of common eigenstates of $\hat{K}^{2}$ and $\hat{K}_{0}$: 
\begin{equation}
\hat{K}^{2}|k,\mu \rangle = k(k-1)|k,\mu \rangle \,,  \qquad \quad 
\hat{K}_{0}|k,\mu \rangle =\mu |k,\mu \rangle \, .   
\end{equation}
The action of the ladder operators $\hat{K}_{\pm }$ on this basis $\{|k,\mu \rangle \}$ is 
\begin{equation}
\hat{K}_{\pm }|k,\mu \rangle =\sqrt{(\mu \pm k)(\mu \mp k\pm 1)}|k,\mu \pm 1\rangle \,,
\end{equation}
and then we can write (making $\mu =k+m$, which is a usual notation) 
\begin{equation}
|k,k+m\rangle =\sqrt{\frac{\Gamma (2k)}{m!\Gamma (2k+m)}}\hat{K}_{+}^{m}|k,k\rangle \,.  
\label{basis}
\end{equation}

The state $|k,k\rangle $ then serves as our fiducial state, as it is annihilated by $\hat{K}_{-}$: $\hat{K}_{-}|k,k\rangle =0$. The isotropy subgroup $H$ of this state; i.e., the $U(1)$ subgroup of SU(1,1) that leaves  $|k,k\rangle $ invariant, up to a phase, is generated by exponentiating $\hat{K}_{0}$. The resulting manifold $\mathbb{H}_{2} ^{+} = $ SU(1,1)/U(1) is the upper sheet of the unit two-sheeted hyperboloid, the phase space associated with this symmetry~\cite{Hasebe:2019aa}.

The coherent states are constructed in the standard way as orbits of the fiducial vector $|k,k\rangle $~\cite{Perelomov:1972aa}; that is 
\begin{equation}
|\zeta \rangle =\hat{S}(\zeta )|k,k\rangle \,,
\end{equation}
where the displacement operator is 
\begin{equation}
	\op{S}(\zeta ) = \exp \left [ \tfrac{1}{2} \tau (\ee^{- i \phi} \op{K}_{+}  - \ee^{i \phi} \op{K}_{-} ) \right ] \, , \qquad
	0 \leq \tau < \infty \, ,  \quad  0 \leq \phi < 2 \pi \, .
	\label{CS}
\end{equation}
The coordinates $(\tau ,\phi )$ are the radial hyperbolic coordinate and the azimuthal angle on the upper sheet $\mathbb{H}_{2}^{+}$ of the unit hyperboloid, whereas 
\begin{equation}
\zeta =\tanh (\tau /2) \ee^{-i\phi }
\end{equation} 
is the stereographic projection of the point $(\tau ,\phi )$ from the south pole onto the unit disk $\mathbb{D}$. In other words, there is a one-to-one correspondence between the hyperbolic Bloch vector 
\begin{equation}
\mathbf{n}=(\cosh \tau ,\sinh \tau \cos \phi ,\sinh \tau \sin \phi )^{\top}\,,
\end{equation}
and points $\zeta $ in the unit disk $\mathbb{D}$.

The coherent states can be expanded in the basis $|k,k+m\rangle $ as 
\begin{equation}
|\zeta \rangle =(1-|\zeta |^{2})^{k}\sum_{m=0}^{\infty }\sqrt{\frac{\Gamma (m+2k)}{m!\Gamma (2k)}}
\zeta^{m}|k,k+m\rangle \,.  
\label{cs11}
\end{equation}
They resolve the identity 
\begin{equation}
\hat{\openone}=\frac{2k-1}{4\pi }  \int \dd\mu (\zeta )\,|\zeta \rangle \langle \zeta |\,,  
\label{Nk}
\end{equation}
where the invariant measure is given by 
\begin{equation}
\dd \mu (\zeta )=\frac{4\dd^{2}\zeta }{(1-|\zeta |^{2})^{2}}=\sinh \tau \dd\tau \dd\phi \,.  
\label{eq:invmeas}
\end{equation}
These {SU}(1,1) coherent states are not orthogonal; their overlap in the discrete irrep $k$ is given by 
\begin{equation}
|\langle \zeta |\zeta ^{\prime }\rangle |^{2}=
\left( \frac{1+\mathbf{n}\cdot \mathbf{n}^{\prime }}{2}\right)^{-2k},
\end{equation}
where $\mathbf{n}\cdot \mathbf{n}^{\prime }$ is a pseudo-scalar product of hyperbolic Bloch vectors 
\begin{equation}
\mathbf{n}\cdot \mathbf{n}^{\prime }=\cosh \tau \cosh \tau^{\prime } -\cos (\phi -\phi^{\prime })
\sinh \tau \sinh \tau ^{\prime }\equiv \cosh \xi \,.
\label{nn'}
\end{equation}

\section{Harmonic functions on the hyperboloid}

An essential ingredient in the theory is the notion of harmonic functions, which are the eigenfunctions of the Laplace-Beltrami operator in the corresponding phase space~\cite{Brif:1996oj}. In our case, this operator on the upper sheet $\mathbb{H}_{2}^{+}$ reads 
\begin{equation}
\mathcal{L}^{2}=\frac{\partial ^{2}}{\partial \tau ^{2}}+\coth \tau 
\frac{\partial }{\partial \tau }+\frac{1}{\sinh ^{2}\tau }\frac{\partial ^{2}}{\partial \varphi ^{2}}\,,
\end{equation}
and can be seen as a differential realization of the Casimir operator $\hat{K}_{0}^{2}-\hat{K}_{1}^{2}-\hat{K}_{2}^{2}$ in terms of the vector fields 
\begin{eqnarray}
& \hat{K}_{0}\mapsto -i\partial _{\phi }\,,&  \notag \\
&& \\
&\hat{K}_{1}\mapsto i\sin \phi \,\partial _{\tau }+i\cos \phi \coth \tau\,\partial _{\phi }\,,\qquad 
\hat{K}_{2}\mapsto -i\cos \phi \,\partial_{\tau }+i\sin \phi \coth \tau \,\partial _{\phi }\,.&  \notag
\end{eqnarray}
that satisfy the commutation relation \eqref{eq:CR1} on the hyperboloid and the transversality condition $\mathbf{n}\cdot \mathbf{K}=0$.

The corresponding eigenvalue equation is  written as 
\begin{equation}
\mathcal{L}^{2}u_{\lambda m}(\zeta )=-\left( \lambda ^{2}+\tfrac{1}{4} \right) u_{\lambda m}(\zeta )\,,  
\label{L EV}
\end{equation}
and the normalized solutions are 
\begin{equation}
u_{\lambda m}(\zeta )=(-1)^{m}\frac{\Gamma \left( \frac{1}{2}+i\lambda \right) }{\Gamma \left( \frac{1}{2}+i\lambda +m\right)} P_{-1/2+i\lambda}^{m}(\cosh \tau)\ee^{im\phi} \, ,
\end{equation}
where $P_{\ell }^{m}$ is an associated Legendre function. The functions $P_{-\frac{1}{2}+i\lambda }^{m}(z)$ are known as Mehler (or conic) functions~\cite{Vilenkin:1991aa} and they are related to the principal continuous series of SU(1,1).  Actually, $\lambda  \tanh(\pi \lambda) \dd \lambda$ is precisely the Plancherel measure for that series. The functions $u_{\lambda m}(\zeta )$ play here the same structural role that spherical harmonics play on the Bloch sphere in the SU(2) setting.

\begin{figure}[t]
\begin{center}
\includegraphics[width=\columnwidth]{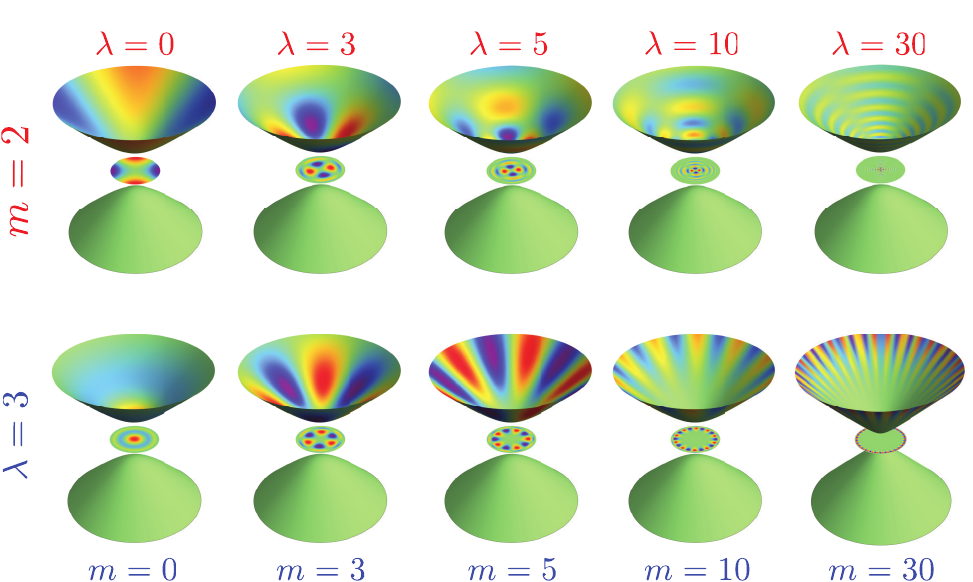}
\end{center}
\caption{Density plots of the real part of the harmonic functions $u_{\lambda m} (\protect\zeta)$ on the upper sheet of the hyperboloid $\mathbb{H}_{2}^{+}$, together with their associated distributions on the unit disk  $\mathbb{D}$ obtained via stereographic projection from the south pole. The corresponding indexes are indicated in each plot.}
\label{fig:uml}
\end{figure}

The continuous index $\lambda \geq 0$ governs the radial oscillation frequency, while $m\in \mathbb{Z}$ controls the azimuthal structure: as illustrated in Fig.~\ref{fig:uml}, each function $u_{\lambda m}(\zeta )$ exhibits $2|m|$ angular lobes, reflecting a growing capacity to resolve finer features on the hyperboloid as $|m|$ increases. This is the hyperbolic counterpart of the familiar angular resolution hierarchy of spherical harmonics.

The functions $u_{\lambda m}(\zeta )$ satisfy the following orthogonality relations 
\begin{equation}
\begin{aligned} 
	\sum_{m=-\infty}^{\infty} \int_0^\infty  \dd\lambda \, \lambda \tanh (\pi \lambda) \, 
	u_{\lambda m} (\zeta) u_{\lambda m}^{\ast} (\zeta^{\prime})
& = 2 \pi \delta (\zeta, \zeta^{\prime}) \, , \\ \lambda \tanh (\pi \lambda)
\int \dd\mu(\zeta) \, u_{\lambda m} (\zeta) u_{\lambda^{\prime} m^{\prime}}^{\ast} (\zeta) & 
= 2 \pi \delta_{m m^{\prime}} \delta (\lambda - \lambda^{\prime}) \, , \\ 
\end{aligned}
\end{equation}
where 
\begin{equation}
\delta (\zeta ,\zeta ^{\prime })=\delta (\cosh \tau -\cosh \tau ^{\prime })\delta (\phi -\phi ^{\prime })\,.
\end{equation}

The convolution of $u_{\lambda m}(\zeta )$ on the index $m$ defines the zonal functions on $\mathbb{H}_{2}$, which are precisely the conic functions: 
\begin{equation}
\sum_{m=-\infty }^{\infty }u_{\lambda m}(\zeta )u_{\lambda m}^{\ast }(\zeta^{\prime })= P_{-\frac{1}{2}+i\lambda }(\cosh \xi )\,,
\end{equation}
with $\xi $ defined in \eqref{nn'}. The important result for our purposes here is that any square-integrable function $f(\tau ,\phi )\equiv f(\zeta )$ on $\mathbb{H}_{2}^{+}$ can be expanded in terms of the harmonic functions as 
\begin{equation}
f(\zeta )=\frac{1}{2\pi }\sum_{m=-\infty }^{\infty }\int_0^\infty  \dd \lambda \,\lambda
\tanh (\pi \lambda )\,f_{\lambda m}u_{\lambda m}(\zeta )\,,
\end{equation}
with coefficients 
\begin{equation}
f_{\lambda m}=\int \dd\mu (\zeta )\,f(\zeta )u_{\lambda m}^{\ast }(\zeta )\,,
\end{equation}
which is known as the Mehler-Fock transform~\cite{Mehler:1881aa,Fock:1943aa}. In this way, any phase-space distribution is decomposed into modes of definite symmetry, and will be the key tool for the multipole analysis developed in Sec.~\ref{sec:multipoles}.

\section{Hyperbolic probability quasidistributions}

The SU(1,1) coherent states provide a natural starting point for constructing quasiprobability distributions on $\mathbb{H}_{2}$. The Husimi $Q$ and Glauber-Sudarshan $P$ distributions are defined, in direct analogy with the continuous-variable case, by 
\begin{equation}
\begin{aligned} 
	Q(\zeta ) & = \langle \zeta |\hat{\varrho}|\zeta \rangle , \\ 
	\\ 
	\hat{\varrho} & = \displaystyle \frac{2k-1}{4\pi}\int \dd\mu (\zeta )
P(\zeta) \; |\zeta \rangle \langle \zeta | \, ,
 \end{aligned}
\end{equation}
where $\hat{\varrho}$ is the density matrix.

A crucial point for what follows is the observation that these two distributions are related by a positively defined, bounded SU(1,1)-invariant spectral operator as~\cite{Klimov:2021aa} 
\begin{equation}
Q(\zeta )=\mathbf{\Phi} _{k} (\mathcal{L}^{2}) \,P(\zeta ),
\end{equation}
where 
\begin{equation}
\mathbf{\Phi}_{k}  (\mathcal{L}^{2})  = - \frac{\pi \mathcal{L}^{2}}{\cos \left( \pi /2\sqrt{1+4\mathcal{L}^{2}}\right)} \prod_{m=1}^{2k-2}\left[ 1-\frac{\mathcal{L}^{2}}{m(m+1)}\right]   
	\label{Piop}
\end{equation}
is a smoothing operator mapping the delta-function over the hyperboloid into the coherent-state overlap 
\begin{equation}
\frac{4\pi }{2k-1}\mathbf{\Phi }_{k}( \mathcal{L}^{2}) \, \delta (\zeta ,\zeta^{\prime }) = 
|\langle \zeta |\zeta ^{\prime }\rangle |^{2} \, .
\end{equation}

This smoothing is the key that allows us to interpolate continuously between all quasidistributions while maintaining full covariance. More generally, the complete family of $s$-parametrized distributions, satisfying all Stratonovich-Weyl conditions (normalization, traciality, and covariance), is related by the integral transform 
\begin{equation}
W^{(s)}(\zeta )=\mathbf{\Phi}_{k}^{{\small {\frac{1}{2}}(s^{\prime }-s)}}(\mathcal{L}^{2})\;W^{(s^{\prime })}(\zeta )=\frac{1}{2\pi }\int \dd\mu (\zeta^{\prime })\;g_{{\small {\frac{1}{2}}(s^{\prime }-s)}}(\zeta ^{\prime-1}\zeta )\;W^{(s^{\prime })}(\zeta ^{\prime })\,,  
\label{exp}
\end{equation}
where the kernel 
\begin{equation}
g_{{t}}(\zeta ^{\prime -1}\zeta )=\int_0^\infty \dd \lambda \,\lambda \tanh (\pi \lambda)\Phi _{k}^{t}(\lambda )
\,P_{-\frac{1}{2}+i\lambda}(\cosh \xi )=g_{t} (\cosh \xi )\, ,  
	\label{gt}
\end{equation}
is expressed through the spectral symbol of $\mathbf{\Phi }_{k}$, which satisfies the condition
\begin{equation}
\mathbf{\Phi}_{k}( \mathcal{L}^{2}) \, u_{\lambda m}(\zeta)=
\Phi _{k}(\lambda )u_{\lambda m}(\zeta ) \, ,  
\label{phi}
\end{equation}
and  reads
\begin{equation}
	\Phi_{k}(\lambda) = \frac{(2k-1)|\Gamma(2k-1/2+i\lambda)|^{2}}{\Gamma^{2}(2k)} \, .
	\label{phi2}
\end{equation}
It is obtained from~\eqref{Piop} by the substitution $\mathcal{L}^{2}\mapsto - (\lambda ^{2}+1/4)$. We also note that 
\begin{equation}
P_{-\frac{1}{2}+i\lambda }(\zeta ^{\prime -1}\zeta )=\sum_{m=-\infty}^{\infty} u_{\lambda m}(\zeta )
u_{\lambda m}^{\ast }(\zeta ^{\prime })=P_{-\frac{1}{2}+i\lambda }(\cosh \xi )\,.
\end{equation}
The special cases $s=-1,0,+1$ recover the $Q$, $W$, and $P$ quasidistributions, respectively.

We now come to the central result of this section. For a coherent state $|\zeta _{0}\rangle $, the $Q$-function is simply the squared overlap 
\begin{equation}
Q_{\zeta _{0}}(\zeta )= |\langle \zeta |\zeta _{0}\rangle |^{2}= \left( \frac{1+\mathbf{n}\cdot \mathbf{n}_{0}}{2}\right) ^{-2k}\,,
\end{equation}
while the $P$-function is a delta function on the hyperboloid 
\begin{equation}
P_{\zeta _{0}}(\zeta )=\frac{4\pi }{2k-1}\delta (\zeta ,\zeta _{0})=
\frac{4\pi }{2k-1}\delta (\cosh \tau -\cosh \tau _{0})\delta (\phi -\phi _{0})\, .
\label{PCS}
\end{equation}
In the Appendix we prove that the integration kernel in \eqref{exp} is positive,  $g_{t}(\cosh xi)>0$,  $\I \xi=0$, for $t >0$. Therefore, every $s$-parametrized quasidistribution, $-1 \leq s<1,$ of a coherent state is strictly positive: 
\begin{equation}
W_{\zeta _{0}}^{(s)}(\zeta )=\frac{4\pi }{2k-1}\mathbf{\Phi}_{k}^{\small{\frac{1}{2}}(1-s)} (\mathcal{L}^{2})\,
\delta (\zeta ,\zeta _{0})=\frac{2}{2k-1}{g}_{{\small {\frac{1}{2}}(1-s)}}(\zeta_{0}^{-1}  \zeta )\,.
\end{equation}

In particular, the Wigner function $W(\zeta ) \equiv W^{(0)}(\zeta )$ for a coherent state is 
\begin{equation}
{W}_{\zeta _{0}}(\zeta )=\frac{2}{2k-1}{g}_{1/2}(\zeta_{0}^{-1}  \zeta ) >0 \, . 
 \label{WCS}
\end{equation}%
This is a qualitatively stronger result than what holds for SU(2) or E(2), where coherent states exhibit Wigner negativity. Here, by contrast, the hyperbolic geometry of $\mathbb{H}_{2}^{+}$ ensures that all classically admissible states forming a convex hull of coherent states, 
\begin{equation}
\hat{\varrho}=\frac{2k-1}{4\pi } \int_{\mathbb{H}_{2}^{+}} \dd\mu (\zeta^{\prime})\,P_{\varrho }(\zeta^{\prime }) \, |\zeta^{\prime }\rangle \langle \zeta^{\prime }|\, ,\qquad P_{\varrho }(\zeta^{\prime })\geq 0 \, ,
 \label{positiveP}
\end{equation}
have a nonnegative Wigner function: 
\begin{equation}
W_{\varrho }(\zeta )=\frac{2k-1}{4\pi }\int_{\mathbb{H}_{2}^{+}} \dd\mu (\zeta^{\prime })\,
P_{\varrho }(\zeta ^{\prime }) W_{\zeta^{\prime }}(\zeta )\geq 0 \, .
\label{positivePiW}
\end{equation}

Any negativity in $W$ is therefore an unambiguous signature of genuinely quantum behavior, and can be quantified by the negativity volume 
\begin{equation}
\mathcal{N}=\frac{2k-1}{4\pi }\int \dd\mu (\zeta )|W_{\varrho } ( \zeta ) |-1\,.
\end{equation}

A complementary perspective on quantumness is provided by the zeros of the Husimi $Q$. They constitute the SU(1,1) analogue of the Majorana constellation~\cite{Majorana:1932aa}. From the relation 
\begin{equation}
Q(\zeta )=\frac{1}{2\pi } \int \dd\mu (\zeta^{\prime })\,\,g_{1/2}(\zeta^{\prime -1} \zeta) \, W(\zeta ^{\prime }) \, ,
\end{equation}%
it follows that the $Q$ function can vanish only if $W$ has negative regions, giving a sufficient witness of Wigner negativity.

It is important to stress that the converse implication does not hold: the absence of zeros of the $Q$-function does not guarantee the nonnegativity of the corresponding Wigner function. A simple example is provided by the Barut-Girardello coherent state~\cite{Barut:1971yq} , which is the eigenstate of $\hat{K}_{-}$ operator: $ \hat{K}_{-}|k, z \rangle _{\mathrm{BG}}= z  |k, z \rangle _{\mathrm{BG}}$. Expanded in the basis \eqref{basis} it reads 
\begin{equation}
|k, z \rangle _{\mathrm{BG}}= N_{k,z} \sum_{n=0}^{\infty }\frac{z^{n}}{\sqrt{n!\,\Gamma (n+2k)}}|k,k+n\rangle \, ,
\qquad 
N_{k,z} = \frac{|z|^{k-1/2}}{\sqrt{I_{2k-1}(2 |z|)}} \, .
\label{BG}
\end{equation}
Its $Q$-function is related to that of the coherent state \eqref{cs11}  centered at the origin, by an exponential factor
\begin{equation}
Q_{k, z} (\zeta ) = N_{k,z}  (1-|\zeta |^{2})^{2k}\, \exp [ 2\R (z^{\ast} \zeta ) ] >0 \, , 
 \label{QBG}
\end{equation}
 so $Q_{k, z}$ has no zeros in  the Poincar\'e disk. 
 
 The corresponding Wigner function can be formally represented as
\begin{equation}
W_{k, z} (\zeta )=N_{k,z} \exp  \{  2 \R  [ z^{\ast }  \; \zeta_{k}(\mathcal{L}^{2}) ]  \} \, W_{0}(\zeta ),
\end{equation}%
where  $\zeta _{k}(\mathcal{L}^{2})=\mathbf{\Phi }_{k}^{-1/2}(\mathcal{L}^{2}) \, \zeta \, \mathbf{\Phi}_{k}^{1/2}(\mathcal{L}^{2})$ is a nonlocal operator that contains derivatives of arbitrarily high order.  Its
explicit form can be found analytically, though it is cumbersome and we do not reproduce it here.  Crucially, the action of  the exponential  on the coherent state Wigner function does not preserve the positivity: this is precisely the mechanism by
which negative regions appear in $W_{k,z}(\zeta)$, with the deepest minimum located along the same direction as the global maximum, i.e., on the ray $\phi=\arg z$~\cite{Baltazar:2025aa}.

To see this explicitly, take $\zeta$  real for simplicity and expand along the direction of maximum negativity in the asymptotic limit $k\gg1$:
\begin{equation}
	\zeta_{k}(\mathcal{L}^{2})\approx
	r-\frac{\varepsilon}{4}(1-r^{2})^{2}\partial_{r}
	-\frac{\varepsilon^{2}}{16}r(1-r^{2})^{3}\partial_{r}^{2}
	-\frac{\varepsilon^{3}}{64}(1-r^{2})^{4}(1+r^{2})\partial_{r}^{3}
	+O(\varepsilon^{4}),
	\label{zetaL}
\end{equation}
where $r=\tanh(\tau/2)$ and $\varepsilon=(2k-1)^{-1}$. Each order in this expansion plays a distinct role. The first-derivative term generates a nonlinear deformation together with a position-dependent rescaling, and by itself preserves positivity. The second-order term acts as a back-diffusion, while the third-order derivative generates an Airy-type kernel whose sign changes. It is this combination that turns the positive, approximate Wigner function at the origin,
\begin{equation}	
	W_{0}(\zeta)\approx 2\cosh^{-2k} (\tau)  \approx 2\ee^{-4kr^{2}},
\end{equation}
into an oscillatory distribution with negative regions once the exponential operator is applied. 

The rank of the constellation thus provides a sufficient, geometry-adapted signature of quantumness that is, by construction, covariant under SU(1,1) transformations. Together, Wigner negativity and stellar rank offer two complementary and operationally meaningful windows onto the quantum structure of SU(1,1) states. However, since the zero set does not capture zero-free factor, it does not determine unambiguously the complete nonclassical structure of the state.

\section{Examples}

To illustrate the physical content of the formalism developed above, we examine three families of states that are representative of qualitatively different quantum behaviors: generic pure states, for which the stellar function provides a unified description of the $Q$-function zero structure; basis states, which carry a definite excitation number; and cat states, which are coherent superpositions and therefore inherently nonclassical.

For a generic pure state $|\psi \rangle =\sum_{m=0}^{\infty }\psi _{m}|k,k+m\rangle $, the $Q$-function admits the factored representation 
\begin{equation}
Q_{\psi }(\zeta )=(1-|\zeta |^{2})^{2k}|f_{\psi }(\zeta ^{\ast })|^{2},
\end{equation}
in terms of the stellar function 
\begin{equation}
f_{\psi }(\zeta )=\sum_{m=0}^{\infty }\left[ \frac{\Gamma (m+2k)}{m!\Gamma (2k)}\right] ^{1/2}\psi _{m}\zeta ^{m}\,.
\end{equation}%
The zeros of $Q$ are precisely the zeros of $f_{\psi }$ in the unit disk and, according to our previous discussion, they signal the presence of Wigner negativity. The stellar function thus provides a covariant information about dark points (unseen by coherent states), however, in  contrast to the  SU(2) symmetry, does not determine the full state.

For a basis state $|k,k+1\rangle $, the $Q$ function takes the form 
\begin{equation}
Q_{1}(\zeta )=2k(1-|\zeta |^{2})^{2k}|\zeta |^{2},
\end{equation}
and has a single (degenerate) zero at the origin, $\tau =0$. This degeneracy is a direct consequence of the rotational symmetry of the state about the $K_{0}$ axis.

The Wigner function, which inherits this rotational invariance, reads 
\begin{equation}
W_{1}(\zeta )=\frac{1}{(2k-1)k}\left( 2k+\frac{\partial ^{2}}{\partial \tau
^{2}}+\coth \tau \frac{\partial }{\partial \tau }\right) {g}_{1/2}(\cosh \tau ) \,.
\end{equation}
This expression exhibits oscillatory behavior with pronounced negative regions. The negativity grows with the excitation number $n$ \cite{Klimov:2021aa}, reflecting the increasing departure from the classical phase-space portrait as the state climbs the ladder of the discrete series. This behavior is the SU(1,1) analogue of the well-known oscillations in the Wigner function of Fock states in quantum optics. 

\begin{figure}[t]
\begin{center}
\includegraphics[width=.65\columnwidth]{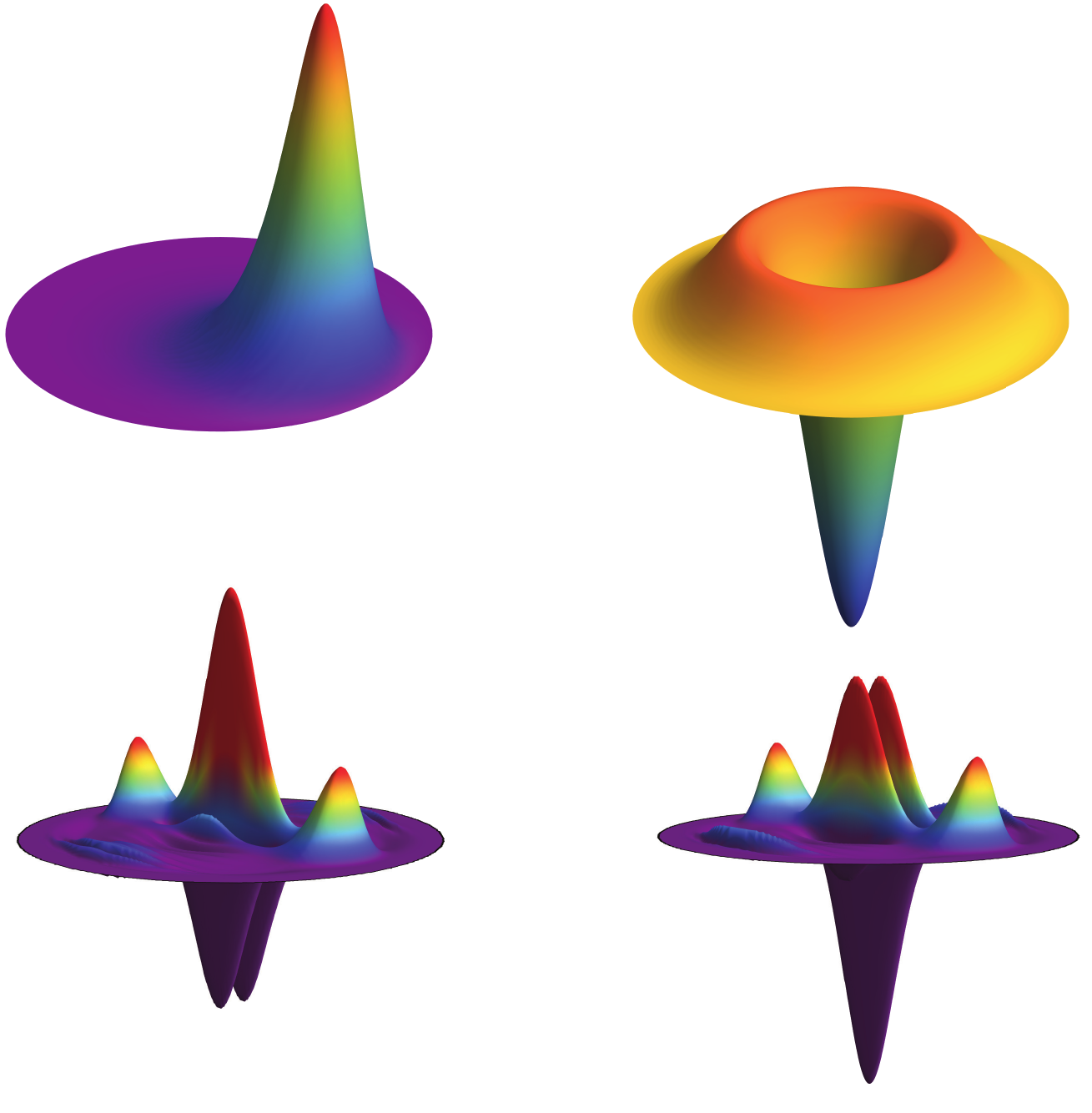}
\end{center}
\caption{Plots of the Wigner function on the unit disk for the states discussed in the text. Upper panel: coherent state (left) and basis state $|k, \protect\mu \rangle$. Lower panel: even cat state (left) and odd cat
state (right).}
\label{fig:Wigv}
\end{figure}

A richer nonclassical structure emerges for even and odd coherent superpositions, 
\begin{equation}
|\Psi \rangle =\frac{N_{\pm }}{\sqrt{2}}(|\zeta _{0}\rangle \pm |-\zeta_{0}\rangle )\,,  
\label{cats}
\end{equation}
where $\zeta _{0}=\tanh (\tau _{0}/2)$ and the normalization constant is $N_{\pm }=(1\pm \cosh ^{-2k}\tau _{0})^{-1/2}$. The Wigner function reads 
\begin{equation}
{W}_{\Psi }(\tau ,\phi )=\frac{N_{\pm }^{2}}{2k-1}\left[ g_{1/2}(\cosh \xi_{+})+
g_{1/2}(\cosh \xi _{-})\pm \frac{2}{\cosh ^{2k}(\tau _{0})}
\R  g_{1/2}(z(\tau ,\phi ))\right] ,  \label{Wcat}
\end{equation}
where the arguments 
\begin{eqnarray}
\cosh \xi _{\pm } &=&\cosh \tau \cosh \tau _{0}\mp \cos \phi \sinh \tau
\sinh \tau _{0}\,,  \notag \\
&& \\
z(\tau ,\phi ) &=&\frac{\cosh \tau -i\sinh \tau _{0}\sinh \tau \sin \phi }{%
\cosh \tau _{0}}\,,  \notag
\end{eqnarray}
encode the hyperbolic distances from the two component coherent states, and the interference geometry between them, respectively. 

The first two terms in (\ref{Wcat}) are the incoherent contributions of each component and are individually positive; the third term is the quantum interference, which can take negative values and is responsible for the quantum character of the superposition. Its magnitude is controlled by the overlap $\cosh ^{-2k}(\tau _{0})$: as the two components are pushed further apart on the hyperboloid ($\tau_{0}\rightarrow \infty $), the interference term is exponentially suppressed and the cat state approaches a classical
mixture.

The corresponding stellar function takes the closed form 
\begin{equation}
f_{\Psi }(\zeta )=(1-|\zeta _{0}|^{2})^{k}\left[ (1-\zeta \zeta_{0})^{-2k}\pm (1+\zeta \zeta _{0})^{-2k}\right] ,
\end{equation}
with zeros for the odd state in the unit disk located at 
\begin{equation}
\zeta \zeta _{0}=\frac{\ee^{i\pi n/k}-1}{\ee^{i\pi n/k}+1}
\end{equation}
with an analogous expression for the even case,
\begin{equation}
\zeta \zeta _{0}=\frac{\ee^{i\pi (2n+1)/2k}-1}{\ee^{i\pi (2n+1)/2k}+1},
\end{equation}
where $n =0, 1 ,\ldots$ and  the solutions satisfying $|\zeta |<1$ correspond to zeros inside the Poincar\'e disk.

Figure~\ref{fig:Wigv} displays the Wigner function for the representative states discussed in this section, illustrating the progression from the strictly positive distribution of coherent states to the increasingly structured negative regions of basis and cat states.

\section{SU(1,1) multipoles}
\label{sec:multipoles}

As discussed in Ref.~\cite{Baltazar:2025aa} one can define tensor operators as 
\begin{equation}
\hat{T}_{\lambda m}^{(k)}=\sqrt{\frac{2k-1}{4\pi }}\Phi _{k}^{- \small{\frac{1}{2}}}(\lambda )\int \dd\mu (\zeta ) | \,  
\zeta \rangle \langle \zeta |u_{\lambda m}(\zeta )\,.  \label{T}
\end{equation}
They form an orthonormal covariant operational basis 
\begin{equation}
\Tr [ \hat{T}_{\lambda m}^{(k)} \, \hat{T}_{\lambda^{\prime }m^{\prime}}^{(k)\dagger}]=
2\pi \delta _{mm^{\prime }} \frac{\delta (\lambda -\lambda ^{\prime })}{\lambda \tanh {(\pi \lambda )}},
\label{TrTT}
\end{equation}
that transforms according to the principal continuous representation of the group. They are the analog of irreducible tensors in SU(2), in terms of which one defines the state multipoles~\cite{Bluhm:1995}.

As for SU(2), any density operator $\hat{\varrho}$ can be expanded in this irreducible tensor basis. The corresponding coefficients, 
\begin{equation}
\varrho _{\lambda m}^{(k)}=\Tr [  \hat{\varrho} \,\hat{T}_{\lambda m}^{(k)\dagger }] =
\sqrt{\frac{2k-1}{4\pi }}\Phi _{k}^{-\small{\frac{1}{2}}}(\lambda )\int \dd\mu (\zeta ) \, \langle \zeta |\hat{\varrho}|\zeta \rangle
\,  u_{\lambda m}^{\ast }(\zeta )\,,
\end{equation}
are the multipoles. These objects have been used to make an accurate assessment of the quantumness of the state~\cite{Goldberg:2020aa}. Accordingly, it is natural to define the SU(1,1)-invariant multipole strength of order $\lambda $ as 
\begin{equation}
q_{\varrho }(k,\lambda )=\sum_{m=-\infty }^{\infty }|\varrho^{(k)}_{\lambda
m}|^{2} \, .
\end{equation}

\begin{figure}[t]
\begin{center}
\includegraphics[width=.95\columnwidth]{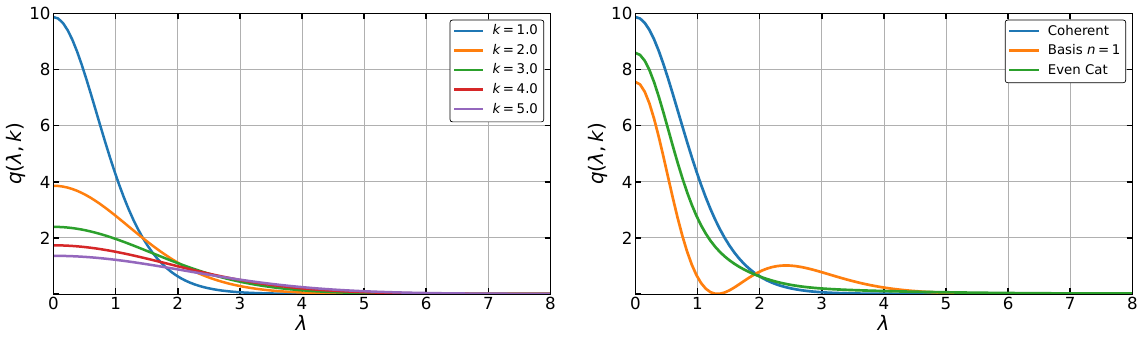}
\end{center}
\caption{Multipole norm as a function of the order $\lambda$ for (left panel) a coherent state for different values of the Bargmann parameter $k$ and (right panel) the different states considered in the text all of them with $k=1$. }
\label{fig:multi}
\end{figure}

Let us examine how these multipoles look for the examples we were considering thus far. For a coherent state $|\zeta _{0}\rangle $ we have 
 \begin{equation}
	\varrho _{\lambda m}^{(k)}(\zeta _{0})=\sqrt{\frac{4\pi }{2k-1}}\Phi _{k}^{\small{\frac{1}{2}}}(\lambda )u_{\lambda m}^{\ast }(\zeta _{0})\,,
\end{equation}
so that 
\begin{equation}
q_{\mathrm{CS}}(k,\lambda )=\frac{4\pi }{2k-1}\Phi _{k}(\lambda )=4\pi \frac{|\Gamma
(2k-\tfrac{1}{2}+i\lambda )|^{2}}{\Gamma ^{2}(2k)}\,.
\end{equation}%
This function is represented in Fig.~\ref{fig:multi} for several values of the Bargmann index $k$. It decreases monotonically with $\lambda $, indicating that, as expected for a classical state, only the lowest-order multipoles contribute appreciably. Moreover, the larger the value of $k$, the smaller the overall multipole weight.

For $P$-positive states  as in \eqref{positiveP},  we have 
\begin{equation}
\bm{\varrho }_{\lambda }^{(k)}=\frac{2k-1}{4\pi }\int_{\mathbb{H}_{2}^{+}} \dd\mu (\zeta^{\prime })\,P_{\varrho }(\zeta^{\prime}) \bm{\varrho }_{\lambda }^{(k)}(\zeta ^{\prime }) \, ,
\end{equation}
where  $[\bm{\varrho }_{\lambda }^{(k)}]_{m}=\varrho _{\lambda m}^{(k)}$ is a vectorized form of the $\lambda$th multipole. Then we have
\begin{eqnarray}
\sqrt{q_{\varrho}(k,\lambda )} & = &\frac{2k-1}{4\pi } \left \Vert \int_{\mathbb{H}_{2}^{+}} \dd
\mu (\zeta^{\prime })\  ,P_{\varrho }(\zeta^{\prime })\bm{\varrho }_{\lambda }^{(k)}(\zeta^{\prime }) 
\right \Vert \leq 
\frac{2k-1}{4\pi }\int_{\mathbb{H}_{2}^{+}} \dd\mu (\zeta^{\prime }) \,P_{\varrho }(\zeta^{\prime })
 \Vert \bm{\varrho }_{\lambda}^{(k)}(\zeta^{\prime })  \Vert \nonumber \\
& = & \sqrt{q_{\mathrm{CS}}(k,\lambda )},
\end{eqnarray}
so that
\begin{equation}
q_{\varrho }(k,\lambda ) \leq q_{\mathrm{CS}}(k,\lambda ) \, ,  
\label{qq}
\end{equation}
which can be considered as a necessary condition of classicality. Therefore, the violation of  \eqref{qq} is a sufficient witnes of SU(1,1) quantumness.

\begin{table}[b]
	\caption{Values of the area $\mathcal{A}_k$ for the states considered in the text.}
	\label{Table}
	\centering
	\begin{tabular}{lccccc}
		\hline
		State & $\mathcal{A}_{k=1}$ & $\mathcal{A}_{k=2}$ & $\mathcal{A}_{k=3}$ & $\mathcal{A}_{k=4}$ & $\mathcal{A}_{k=5}$ \\ \hline
		Coherent & 9.86959454 & 6.16849863 & 4.85768610 & 4.13475597 & 3.66073054 \\ 
		Even Cat & 8.01706295 & 4.05507036 & 2.82796851 & 2.28982217 & 1.99278990 \\ 
		Basis $m=1$ & 6.16849308 & 4.24072028 & 3.43940109 & 2.96458112 & 2.63265090
		\\ 
		Odd Cat & 5.95668850 & 3.75684554 & 2.77938358 & 2.28193040 & 1.99150343 \\ 
		\hline
	\end{tabular}
	\label{tab:estados_Ak}
\end{table}

For the basis state $|k,k+n\rangle $, one can use the representation in Ref.~\cite{Baltazar:2025aa} for the tensor operators, a lengthy calculation then yields 
\begin{equation}
q_{n}(k,\lambda )=4\pi \frac{|\Gamma (2k-\tfrac{1}{2}+i\lambda )|^{2}}{\Gamma ^{2}(2k)}\frac{1}{[n!(2k)_{n}]^{2}}
\left\vert S_{n}\left( \lambda^{2};\frac{1}{2};2k-\frac{1}{2},\frac{1}{2}\right) \right\vert ^{2}\,,
\end{equation}
where $(\cdot )_{n}$ is the Pochhammer symbol and $S_{n}$ are the continuous dual Hahn polynomials~\cite{NIST:specfun} that can be equivalently expressed in terms of hypergeometric functions.

Finally, for the cat states the multipoles can also be obtained in closed form, but the resulting expression is long and offers little additional physical insight, so we omit it here. Figure~\ref{fig:multi} shows the multipole strength for all three families of states.

A more quantitative comparison is provided by the multipole area,  
\begin{equation}
\mathcal{A}_{k}=\int_{0}^{\infty }\dd\lambda \, q(k,\lambda )\,,
\end{equation}
which for coherent states takes the form
\begin{equation}
\mathcal{A}_{\mathrm{CS}}=4\pi ^{2}\frac{\Gamma (4k-1)}{2^{4k-1}\Gamma^{2}(2k)}.
\end{equation}
It follows frrm \eqref{qq} that for $P$-positive states \eqref{positiveP} $\mathcal{A}_{\varrho }\leq \mathcal{A}_{\mathrm{CS}}$. 

The results of this integration are given in Table~\ref{tab:estados_Ak}, which complements the visual information of Fig.~\ref{fig:multi}. The coherent state gives the maximal integrated area for every value of $k$, while the cat states show a reduced total weight, more pronounced for the odd cat. The basis state displays intermediate behavior, its relative position depending on the value of~$k$.

\section{Concluding remarks}

We have analyzed classicality  for systems endowed with SU(1,1) dynamical symmetry, in the framework of a phase-space formalism built on the two-sheeted hyperboloid $\mathbb{H}_{2}$ and its stereographic image, the Poincar\'{e} disk. Using the covariant $s$-parametrized maps, connected through a single smoothing operator, we established a strict  positivity for all quasidistributions, and in particular, the Wigner function,  of Perelomov coherent states. 

This is not a technicality: it is a structural consequence of the hyperbolic geometry of $\mathbb{H}_{2}^{+}$, and it stands in marked contrast to the SU(2) case, where even the most classical states of the theory can display Wigner negativity.  Every SU(1,1) classical state, constructed as a convex hull of coherent states, i.e., admitting $P$-positive representation, has non-negative Wigner function. Consequently, the Wigner negativity is a sufficient unambiguous witness of nonclassicality relative to this symmetry-adapted classical set. Moreover, zeros of the Husimi function imply Wigner negativity. The converse is not true: zero-free Husimi distribution does not guarantee a non-negativity of the Wigner function, in contrast to the Heisenberg-Weyl symmetry.

We illustrated the formalism on three representative families of states: basis states and even and odd coherent superpositions, whose interference fringes are governed by the hyperbolic separation between their components and decay exponentially as that separation grows. In parallel, we introduced the SU(1,1) multipole hierarchy, obtained by expanding the density operator on the covariant tensor basis. The multipole strength $q(k,\lambda )$ and its integrated spectral area $\mathcal{A}_{k}$, furnish a second and independent measure of departure from classicality: the coherent state concentrates its weight at low $\lambda $ and maximizes $\mathcal{A}_{k}$
for every $k$, while the cat states redistribute weight toward higher multipoles, an effect most pronounced for the odd superposition. That Wigner negativity and multipole spreading track the same physical distinctions, while probing the state through entirely different lenses, reinforces the central message of this paper: hyperbolic phase space offers a coherent and self-consistent language for quantumness in SU(1,1) systems, one that does not need to borrow criteria designed for flat or spherical geometries. 

These results emphasize that the concept of classicality is symmetry-relative. Perelomov coherent states are natural classical states of the SU(1,1) orbit, even though their boson realizations may regard them as squeezed or entangled (in case of two-mode systems) states according to the Heisenberg-Weyl group.

\section*{Acknowledgments}

This research was funded by the Spanish Agencia Estatal de Investigaci\'{o}n
(Grant PID2021-127781NB-I00), the~Mexican CONAHCyT (Grant CBF2023-2024-50)
and the Chilean Agencia Nacional de Investigaci\'{o}n y Desarrollo
(Postdoctoral Fellowship 74240083).

\appendix
\section{Positivity of the kernel $g_{{t}}(\cosh \xi )$  }

In this Appendix we prove the positivity of $g_{{t}}(\cosh \xi )$ defined in \eqref{gt} for $t>0$. Let us observe that the spectral function \eqref{phi} can be represented as an infinite product
\begin{equation}
\Phi_{k}^{t}(\lambda )= \prod_{n=0}^{\infty }\left( \frac{b_{nk}}{b_{nk}+q}\right) ^{t}, \qquad 
b_{nk}=(n+c)(n+c+1) \, ,  
\label{Phit}
\end{equation}
where $q=\lambda^{2}+{1}/{4}$ and $c=2k-1>0$. Each factor is the Laplace transform of a normalized Gamma density
\begin{equation}
	\left(\frac{b}{b+q}\right)^{t}=\int_{0}^{\infty}\dd u \,\ee^{-qu}\gamma_{bt}(u),
	\qquad
	\gamma_{bt}(u)=\frac{b^{t}}{\Gamma(t)}u^{t-1}\ee^{-bu},
	\qquad
	\int_{0}^{\infty}\dd u\,\gamma_{bt}(u)=1 \, .
\end{equation}
Since the Laplace transform turns convolution into multiplication, the partial product up to any fixed $N$ is itself a Laplace transform
\begin{equation}
	\int_{0}^{\infty}\dd u\,\ee^{-qu}f_{kt}^{(N)}(u)
	=\prod_{n=0}^{N}\left(\frac{b_{nk}}{b_{nk}+q}\right)^{t}
	=F_{N}(q),
	\label{Laplace}
\end{equation}
of the nonnegative, normalized convolution density
\begin{equation}
	f_{kt}^{(N)} =\gamma_{b_{0,kt}}  \ast \gamma_{b_{1,kt}} \ast \cdots \ast \gamma_{b_{N,kt}},
	\qquad
	\int_{0}^{\infty}\dd u\,f_{kt}^{(N)}(u)=1,
	\qquad f_{k,t}^{(N)}(u)\ge 0 \, .
	\label{convol}
\end{equation}
 
 The tail of $-\ln F_N(q)$ is controlled by
 \begin{equation}
 	-\ln F_{N}(q)=t\sum_{n=0}^{N}\ln\!\left(1+\frac{q}{b_{n,k}}\right)
 	\le t\sum_{n=0}^{N}\frac{q}{b_{nk}}
 	=\frac{qt(N+1)}{(2k-1)(N+2k)} < \frac{qt}{2k-1},
 \end{equation}
so the product converges to a strictly positive number for every $q>0$. The same bound is uniform on compact intervals: for $0\le q\le Q$,
\begin{equation}
	t\sum_{n=N+1}^{\infty}\ln\!\left(1+\frac{q}{b_{nk}}\right)
	\le t\sum_{n=N+1}^{\infty}\frac{Q}{b_{nk}}
	=\frac{Qt}{N+2k}\xrightarrow[N\to\infty]{}0.
\end{equation}
Hence $F_N(q)$ converges uniformly on compacts to a limit
\begin{equation}
	F(q)=\lim_{N\to\infty}F_{N}(q)\ge\exp\!\left(-\frac{qt}{2k-1}\right)>0,
	\qquad q>0,
\end{equation}
which is also bounded above by the leading factor, $F(q)\leq [b_{0,k}/(b_{0k}+q)]^{t}$, so $F(q)\to0$ as $q\to\infty$.

Each $F_N(q)$ is completely monotone, $(-1)^{m}\partial_q^m F_N(q)\ge0$ for $m=0,1,\dots$, being a Laplace transform of a nonnegative density, and this property is preserved under the pointwise limit $F_N\to F$. By the
Bernstein--Widder theorem~\cite{Widder:1941aa}, $F$ is therefore itself the Laplace transform of a unique positive Borel measure $\nu_{k,t}(du)$,
\begin{equation}
	F(q)=\int_{0}^{\infty}\nu_{kt}(\dd u)\,\ee^{-qu}=\Phi_{k}^{t}(\lambda).
\end{equation}
Since $F(q)\to0$ as $q\to\infty$, the measure has no atom at the origin, $\nu_{k,t}(\{0\})=0$, and $F(0)=1$ gives the normalization
$\int_0^\infty \nu_{k,t}(\dd u)=1$. Formally, $\nu_{k,t}(du)$ is the limit of the finite convolution measures~\eqref{convol}.

In operator language, this representation reads
\begin{equation}
	\mathbf{\Phi}_{k}^{t}(\mathcal{L}^{2})=\int_{0}^{\infty}\nu_{kt}(\dd u)\,\ee^{u \mathcal{L}^{2}},
\end{equation}
that is, $\mathbf{\Phi}_{k}^{t}(\mathcal{L}^{2})$ is a positive superposition of hyperbolic heat semigroups,
\begin{equation}
	\ee^{u\mathcal{L}^{2}}f(\zeta)=\int \dd\mu(\zeta')\,h_{u}(\zeta,\zeta')f(\zeta'),
\end{equation}
with heat kernel $h_u(\zeta,\zeta')=h_u(\cosh\xi)$, $u>0$,
\begin{eqnarray}
	h_{u}(\cosh\xi) & = & \frac{1}{2\pi}\int \dd\lambda\,\lambda\tanh(\pi\lambda)\, \ee^{-u(\lambda^{2}+1/4)}
	P_{-\frac{1}{2}+i\lambda}(\cosh\xi) \nonumber \\
	& = & \frac{\sqrt{2}\,\ee^{-u/4}}{(4\pi u)^{3/2}}
	\int_{\xi}^{\infty}\dd r\,\frac{r\,\ee^{-r^{2}/(4u)}}{\sqrt{\cosh r-\cosh\xi}},
\end{eqnarray}
normalized and approaching the delta function as $u\to0^+$,
\begin{equation}
	\int \dd\mu(\zeta')\,h_{u}(\zeta,\zeta')=1,
	\qquad
	\lim_{u\to0^{+}}h_{u}(\zeta,\zeta')=\delta(\zeta,\zeta') \, . 
\end{equation}

The heat kernel is strictly positive on $\mathbb{H}_{2}^{+}$: it preserves positivity and strictly improves it for any nonidentically-zero
$f\ge0$, we have $ \ee^{\mathcal{L}^{2}u}f(\zeta)>0$. Moreover, the integral~\eqref{gt} is absolutely convergent, since
$\Phi_k(\lambda)\sim\lambda^{4k-2}\ee^{-\pi|\lambda|}$ for $\lambda\gg1$. Combining
these facts,
\begin{equation}
	\mathbf{\Phi}_k^{t}(\mathcal{L}^{2})\,\delta(\zeta,\zeta_{0})
	=\frac{\sqrt{2}}{(4\pi)^{3/2}} \int_{0}^{\infty}\nu_{kt}(\dd u)\,\frac{\ee^{-u/4}}{u^{3/2}}
	\int_{\xi}^{\infty}\dd r\,\frac{r\,\ee^{-r^{2}/(4u)}}{\sqrt{\cosh r-\cosh\xi}} >0,
	\qquad t>0,
\end{equation}
and therefore
\begin{equation}
	g_{t}(\cosh\xi)=2\pi\int_{0}^{\infty}\nu_{kt}(\dd u)\,h_{u}(\cosh\xi)>0, 
\end{equation} 
as we wanted to demonstrate.


\begin{thebibliography}{62}
	\providecommand{\natexlab}[1]{#1}
	\providecommand{\url}[1]{\texttt{#1}}
	\expandafter\ifx\csname urlstyle\endcsname\relax
	\providecommand{\doi}[1]{doi: #1}\else
	\providecommand{\doi}{doi: \begingroup \urlstyle{rm}\Url}\fi
	
	\bibitem[Schroek(1996)]{Schroek:1996fv}
	F.~E. Schroek.
	\newblock \emph{Quantum Mechanics on Phase Space}.
	\newblock Kluwer, Dordrecht, 1996.
	
	\bibitem[Schleich(2001)]{Schleich:2001hc}
	W.~P. Schleich.
	\newblock \emph{Quantum Optics in Phase Space}.
	\newblock Wiley-VCH, Berlin, 2001.
	
	\bibitem[Zachos et~al.(2005)Zachos, Fairlie, and Curtright]{QMPS:2005mi}
	C.~K. Zachos, D.~B. Fairlie, and T.~L. Curtright, editors.
	\newblock \emph{Quantum Mechanics in Phase Space}.
	\newblock World Scientific, Singapore, 2005.
	
	\bibitem[Weinbub and Ferry(2018)]{Weinbub:2018aa} 
 J.~Weinbub and D.~K. Ferry. \href{https://doi.org/10.1063/1.5046663}  
  {\emph{Recent advances in Wigner function approaches}, 
	Appl. Phys. Rev. \textbf{5}, 041104 (2018).}
	
	\bibitem[Kirillov(2004)]{Kirillov:2004aa}
	A.~A. Kirillov.
	\newblock \emph{Lectures on the Orbit Method}, volume~64 of \emph{Graduate
		Studies in Mathematics}.
	\newblock American Mathematical Society, Providence, 2004.
	
	\bibitem[Perelomov(1986)]{Perelomov:1986kl}
	A.~Perelomov.
	\newblock \emph{Generalized Coherent States and their Applications}.
	\newblock Springer, Berlin, 1986.
	
	\bibitem[Gazeau(2009)]{Gazeau:2009aa}
	J.-P. Gazeau.
	\newblock \emph{{Coherent States in Quantum Physics}}.
	\newblock Wiley, New York, 2009.
	
	\bibitem[Robert and Combescure(2021)]{Robert:2021aa}
	D.~Robert and M.~Combescure.
	\newblock \emph{Coherent States and Applications in Mathematical Physics}.
	\newblock Springer, Cham, second edition, 2021.
	
	\bibitem[Streltsov et~al.(2017)Streltsov, Adesso, and Plenio]{Streltsov:2017aa}
	A. Streltsov, G. Adesso, and M.~B. Plenio. \href{https://doi.org/10.1103/RevModPhys.89.041003}
{\emph{Quantum coherence as a resource},  Rev. Mod. Phys.  \textbf{89}, 041003 (2017).}

	
	\bibitem[Chitambar and Gour(2019)]{Chitambar:2019aa}
	E.~Chitambar and G.~Gour. \href{https://doi.org/10.1103/RevModPhys.91.025001}
	{\emph{Quantum resource theories},  Rev. Mod. Phys. \textbf{91}, 025001 (2019).}

	
	\bibitem[Dodonov(2002)]{Dodonov:2002aa}
	V.~V. Dodonov. \href{10.1088/1464-4266/4/1/201}
	{\emph{`Nonclassical' states in quantum optics: a `squeezed' review of the first 75 years},  J. Opt. B: Quantum Semiclass. Opt. \textbf{4}, R1 (2002).}

	
	\bibitem[Mandel(1979)]{Mandel:1979aa}
	L.~Mandel. \href{https://doi.org/10.1364/OL.4.000205}
	{\emph{Sub-Poissonian photon statistics in resonance fluorescence}, Opt. Lett.  \textbf{4}, 205--207 (1979).}

	
	\bibitem[Glauber(1963)]{Glauber:1963aa}
	R.~J. Glauber.\href{https://doi.org/10.1103/PhysRev.131.2766}
	{\emph{Coherent and incoherent states of the radiation field},  Phys. Rev. \textbf{131}, 2766--2788 (1963).}

	\bibitem[Sudarshan(1963)]{Sudarshan:1963aa}
	E.~C.~G. Sudarshan. \href{https://doi.org/10.1103/PhysRevLett.10.277} 
	{\emph{Equivalence of semiclassical and quantum mechanical descriptions of
	statistical light beams}, Phys. Rev. Lett. \textbf{10}, 277--279 (1963).}

	
	\bibitem[Titulaer and Glauber(1965)]{Titulaer:1965aa}
	U.~M. Titulaer and R.~J. Glauber. \href{https://doi.org/10.1103/PhysRev.140.B676}
	{\emph{Correlation functions for  coherent fields},  Phys. Rev. \textbf{140}, B676--B682 (1965).}
	
	\bibitem[Tan et~al.(2020)Tan, Choi, and Jeong]{Tan:2020aa}
	K.~C. Tan, S.Choi, and H. Jeong. \href{https://doi.org/10.1103/PhysRevLett.124.110404}
	{\emph{Negativity of quasiprobability distributions as a measure of nonclassicality},  Phys. Rev. Lett.  \textbf{124}, 110404 (2020).}

	
	\bibitem[Chabaud et~al.(2020)Chabaud, Markham, and Grosshans]{Chabaud:2020aa}
	U. Chabaud, D. Markham, and F. Grosshans. \href{https://doi.org/10.1103/PhysRevLett.124.063605}
	{\emph{Stellar representation of non-Gaussian quantum states},  Phys. Rev. Lett. \textbf{124},  063605 (2020).}

	
	\bibitem[Husimi(1940)]{Husimi:1940aa}
	K.~Husimi. \href{https://doi.org/10.11429/ppmsj1919.22.4_264}
	{\emph{Some formal properties of the density matrix},  Proc. Phys. Math. Soc. Jpn.  \textbf{22}  264--314 (1940).}
	
	\bibitem[Majorana(1932)]{Majorana:1932aa}
	E.~Majorana. \href{http://dx.doi.org/10.1007/BF02960953}
	{\emph{Atomi orientati in campo magnetico variabile},  Nuovo Cimento  \textbf{9}, 43--50 (1932).}

	
	\bibitem[Kenfack and {\.Z}yczkowski(2004)]{Kenfack:2004lw}
	A.~Kenfack and K.  {\.Z}yczkowski. \href{https://doi.org/10.1088/1464-4266/6/10/003}
   {\emph{ Negativity of the Wigner function as an indicator of non-classicality}, J. Opt. B: Quantum Semiclass. Opt. \textbf{6},  396--404 ( 2004).}
	
	\bibitem[Hudson(1974)]{Hudson:1974kb}
	R.~L. Hudson. \href{https://doi.org/10.1016/0034-4877(74)90007-X}
	{\emph{When is the Wigner quasi-probability density non-negative?},  Rep. Math. Phys. \textbf{6}, 249--252 (1974).}
	
	\bibitem[Yoshikawa et~al.(2013)Yoshikawa, Makino, Kurata, van Loock, and
	Furusawa]{Yoshikawa:2013aa}
	J. Yoshikawa, K. Makino, S. Kurata, P. van Loock, and A.Furusawa. \href{https://doi.org/10.1103/PhysRevX.3.041028}
	{\emph{Creation, storage, and on-demand release of optical quantum states with a negative Wigner function},  Phys. Rev. X \textbf{3}, 041028 (2013).}
	
	
	\bibitem[Fl{\"u}hmann et~al.(2019)Fl{\"u}hmann, Nguyen, Marinelli, Negnevitsky,
	Mehta, and Home]{Fluhmann:2019aa}
	C.~Fl{\"u}hmann, T.~L. Nguyen, M.~Marinelli, V.~Negnevitsky, K.~Mehta, and
	J.~P. Home. \href{https://doi.org/10.1038/s41586-019-0960-6}
	{\emph{Encoding a qubit in a trapped-ion mechanical oscillator},  Nature  \textbf{566}, 513--517 (2019).}

	
	\bibitem[Lu et~al.(2021)Lu, Strandberg, Quijandr{\'\i}a, Johansson,
	Gasparinetti, and Delsing]{Lu:2021aa}
	Y. Lu, I. Strandberg, F. Quijandr{\'\i}a, G. Johansson, S. Gasparinetti, and P. Delsing. 
	\href{https://doi.org/10.1103/PhysRevLett.126.253602}
	{\emph{Propagating Wigner-negative states generated from the steady-state emission of a superconducting qubit},
	Phys. Rev. Lett. \textbf{126}, 253602 (2021).}


	\bibitem[Stratonovich(1956)]{Stratonovich:1956aa}
	R.~L. Stratonovich.  \href{https://jetp.ras.ru/cgi-bin/e/index/e/4/6/p891?a=list}
	{\emph{On distributions in representation space},  JETP \textbf{31}, 1012--1020 (1956).}

	\bibitem[Berezin(1975)]{Berezin:1975aa}
	F.~A. Berezin.\href{https://doi.org/10.1007/BF01609397}
	{\emph{General concept of quantization}, Commun. Math. Phys. \textbf{40}, 153--174 (1975).}
	
	\bibitem[Varilly and Gracia-Bond{\'{\i}}a(1989)]{Varilly:1989bh}
	J.~C. Varilly and J.~M. Gracia-Bond{\'{\i}}a. \href{https://doi.org/10.1016/0003-4916(89)90262-5}
	{\emph{The Moyal representation for spin}, Ann. Phys. \textbf{190}, 107--148 (1989).}

	\bibitem[Klimov et~al.(2017)Klimov, Romero, and de~Guise]{Klimov:2017aa}
	A.~B. Klimov, J.~L. Romero, and H.~de~Guise. \href{http://dx.doi.org/10.1088/1751-8121/50/32/323001}
	{\emph{Generalized su(2) covariant Wigner functions and some of their applications},  J. Phys. A: Math. Theor. \textbf{50}, 323001 (2017).}

	\bibitem[Davis et~al.(2021)Davis, Kumari, Mann, and Ghose]{Davis:2021aa}
	J. Davis, M. Kumari, R.~B. Mann, and S. Ghose. \href{https://doi.org/10.1103/PhysRevResearch.3.033134}
	{\emph{Wigner negativity in spin-{$J$} systems}, Phys. Rev. Res.  \textbf{3}, 033134 (2021).}

	\bibitem[Mukunda(1979)]{Mukunda:1979uq}
	N.~Mukunda. \href{https://doi.org/10.1119/1.11869}
	{\emph{Wigner distribution for angle coordinates in quantum mechanics},  Am. J. Phys. \textbf{47}, 182--187 (1979).}

	\bibitem[Pleba\'{n}ski et~al.(2000)Pleba\'{n}ski, Prazanowski, Tosiek, and
	Turrubiates]{Plebanski:2000fk} 
	J.~F. Pleba\'{n}ski, M.~Prazanowski, J.~Tosiek, and F.~K. Turrubiates. \href{https://www.actaphys.uj.edu.pl/fulltext?series=Reg&vol=31&page=561}
	{\emph{Remarks on deformation quantization on the cylinder},  Acta Phys. Pol. B \textbf{31},  561--587 (2000).}
	
	\bibitem[Rigas et~al.(2011)Rigas, S{\'a}nchez-Soto, Klimov, \v{R}eh\'a\v{c}ek,
	and Hradil]{Rigas:2011by}
	I.~Rigas, L.~L. S{\'a}nchez-Soto, A.~B. Klimov, J.~\v{R}eh\'a\v{c}ek, and
	Z.~Hradil. \href{https://doi.org/10.1016/j.aop.2010.11.016}
   {\emph{Orbital angular momentum in phase space},  Ann. Phys.  \textbf{326}, 426--439 (2011).}

	
	\bibitem[Kastrup(2016)]{Kastrup:2016aa}
	H.~A. Kastrup. \href{https://doi.org/10.1103/PhysRevA.94.062113}
	{\emph{Wigner functions for the pair angle and orbital angular momentum},  Phys. Rev. A \textbf{94}, 062113 (2016).}

	
	\bibitem[Fabre et~al.(2023)Fabre, Klimov, Murenzi, Gazeau, and
	S{\'a}nchez-Soto]{Fabre:2023aa}
	N. Fabre, A.~B. Klimov, R. Murenzi, J.-P. Gazeau, and	L.~L. S{\'a}nchez-Soto. \href{https://doi.org/10.1103/PhysRevResearch.5.L032006}
	{\emph{Majorana stellar representation of twisted photons}, Phys. Rev. Res. \textbf{5}, L032006 (2023).}

	\bibitem[Wodkiewicz and Eberly(1985)]{Wodkiewicz:1985aa}
	K.~Wodkiewicz and J.~H. Eberly. \href{https://doi.org/10.1364/JOSAB.2.000458} 
	{\emph{Coherent states, squeezed fluctuations, and the {SU(2)} and {SU(1,1)}
	groups in quantum-optics applications}, J. Opt. Soc. Am. B  \textbf{2}, 458--466 (1985).}
	
	\bibitem[Gerry(1985)]{Gerry:1985aa}
	C.~C. Gerry. \href{https://doi.org/10.1103/PhysRevA.31.2721} 
	{\emph{Dynamics of {SU(1,1)} coherent states},  Phys. Rev. A \textbf{31},  2721--2723 (1985).}

	\bibitem[Gerry(1991)]{Gerry:1991aa}
	C.~C. Gerry. \href{https://doi.org/10.1364/JOSAB.8.000685}
	{\emph{Correlated two-mode {SU(1, 1)} coherent states: nonclassical properties},  J. Opt. Soc. Am. B \textbf{8}, 685--690 (1991).}

	\bibitem[Gerry and Grobe(1995)]{Gerry:1995kq}
	C.~C. Gerry and R.~Grobe. \href{https://doi.org/10.1103/PhysRevA.51.4123}
	{\emph{Two-mode intelligent {SU(1,1)} states}, Phys. Rev.  A \textbf{51}, 4123--4131 (1995).}	
	
	\bibitem[Gazeau and del Olmo(2023)]{Gazeau:2023aa}
	J.~P. Gazeau and M.~A. del Olmo. \href{https://doi.org/10.1364/JOSAB.484284}
	{\emph{SU(1,1)-displaced coherent states, photon counting, and squeezing},  J. Opt. Soc. Am. B  \textbf{40}, 1083--1091 (2023).}

	\bibitem[del Olmo and Gazeau(2020)]{Olmo:2020aa}
	M.~A. del Olmo and J.~P. Gazeau. \href{https://doi.org/10.1063/1.5128066}
	{\emph{Covariant integral quantization of the unit disk}, J. Math. Phys. \textbf{61},  022101  (2020)}.

	\bibitem[Jing et~al.(2011)Jing, Liu, Zhou, Ou, and Zhang]{Jing:2011aa}
	J.~Jing, C.~Liu, Z.~Zhou, Z.~Y. Ou, and W.~Zhang. \href{https://doi.org/10.1063/1.3606549}
	{\emph{Realization of a nonlinear interferometer with parametric amplifiers},  Appl. Phys. Lett. \textbf{99}, 011110 (2011).}

	\bibitem[Hudelist et~al.(2014)Hudelist, Kong, Liu, Jing, Ou, and
	Zhang]{Hudelist:2014aa}
	F.~Hudelist, J.~Kong, C.~Liu, J.~Jing, Z.~Y. Ou, and W.~Zhang. \href{https://doi.org/10.1038/ncomms4049}
	{\emph{Quantum metrology with parametric amplifier-based photon correlation interferometers},  Nat. Commun. \textbf{5}, 3049 (2014).}

	
	\bibitem[Yurke et~al.(1986)Yurke, McCall, and Klauder]{Yurke:1986yg}
	B.~Yurke, S.~L. McCall, and J.~R. Klauder. \href{https://doi.org/10.1103/PhysRevA.33.4033}
	{\emph{{SU(2)} and {SU(1,1)} interferometers}, Phys. Rev. A \textbf{33}, 4033--4054 (1986).}
	
	\bibitem[Chekhova and Ou(2016)]{Chekhova:2016aa}
	M.~V. Chekhova and Z.~Y. Ou.  \href{https://doi.org/10.1364/AOP.8.000104}
	{\emph{Nonlinear interferometers in quantum optics}  Adv. Opt. Photon. \textbf{8}, 104--155 (2016).}

	\bibitem[Or{\l}owski and W{\'o}dkiewicz(1990)]{Orowski:1990aa}
	A.~Or{\l}owski and K.~W{\'o}dkiewicz. \href{https://doi.org/10.1080/09500349014550361}
  {\emph{On the {SU(1, 1)} phase-space description of reduced and squeezed
	quantum fluctuations},  J. Mod. Opt. \textbf{37},  295--301 (1990).}

	
	\bibitem[Alonso et~al.(2002)Alonso, Pogosyan, and Wolf]{Alonso:2002aa}
	M.~A. Alonso, G.~S. Pogosyan, and K.~B. Wolf. \href{https://doi.org/10.1063/1.1518139}
	{\emph{Wigner functions for curved spaces. I. On hyperboloids},  J. Math. Phys. \textbf{43}, 5857--5871 (2002).}

	\bibitem[de~la Hoz et~al.(2013)de~la Hoz, Klimov, Bj{\"o}rk, Kim, M{\"u}ller,
	Marquardt, Leuchs, and S{\'a}nchez-Soto]{Hoz:2013aa}
	P.~de~la Hoz, A.~B. Klimov, G.~Bj{\"o}rk, Y.~H. Kim, C.~M{\"u}ller, Ch.
	Marquardt, G.~Leuchs, and L.~L. S{\'a}nchez-Soto. \href{https://doi.org/10.1103/PhysRevA.88.063803} 
	{\emph{Multipolar hierarchy of efficient quantum polarization measures},  Phys. Rev. A \textbf{88},  063803 (2013).} 

	
	\bibitem[Goldberg et~al.(2022)Goldberg, Klimov, de~Guise, Leuchs, Agarwal, and
	S{\'a}nchez-Soto]{Goldberg:2022aa}
	A.~Z. Goldberg, A.~B. Klimov, H.~de~Guise, G.~Leuchs, G.~S. Agarwal, and L.~L.
	S{\'a}nchez-Soto. \href{https://opg.optica.org/ol/abstract.cfm?URI=ol-47-3-477}
	{\emph{From polarization multipoles to higher-order coherences},  Opt. Lett. \textbf{47},  477--480 (2022).}

	
	\bibitem[Goldberg et~al.(2020)Goldberg, Klimov, Grassl, Leuchs, and
	S{\'a}nchez-Soto]{Goldberg:2020aa}
	A.~Z. Goldberg, A.~B. Klimov, M.~Grassl, G.~Leuchs, and L.~L. S{\'a}nchez-Soto.  \href{https://doi.org/10.1116/5.0025819}
	{\emph{Extremal quantum states}  AVS Quantum Sci. \textbf{2}, 044701 (2020).}
	
	\bibitem[Goldberg et~al.(2024)Goldberg, Klimov, Leuchs, and
	Sanchez-Soto]{Goldberg:2024aa}
	A.~Z. Goldberg, A.~B. Klimov, G.~Leuchs, and L.~L. Sanchez-Soto. \href{https://doi.org/10.22331/q-2024-05-29-1363}
	{\emph{Covariant operator bases for continuous variables},  Quantum \textbf{8}, 1363 (2024).}

	\bibitem[Bargmann(1947)]{Bargmann:1947fk}
	V.~Bargmann. \href{https://www.jstor.org/stable/1969129?origin=JSTOR-pdf&seq=1#metadata_info_tab_contents}.
	{\emph{Irreducible unitary representations of the {L}orentz group},  Ann. Math. \textbf{48}, 568--640 (1947).}

	
	\bibitem[Hasebe(2019)]{Hasebe:2019aa}
	K.~Hasebe. \href{https://doi.org/10.1088/1751-8121/ab3cda}
	{\emph{Sp(4;$\mathbb{R}$) squeezing for {B}loch four-hyperboloid via the non-compact {H}opf map}, 
  J. Phys. A: Math. Theor.  \textbf{63}, 055303 (2020).}
	
	\bibitem[Perelomov(1972)]{Perelomov:1972aa}
	A. M.~Perelomov. \href{https://doi.org/10.1007/BF01645091}
	{\emph{Coherent states for arbitrary Lie groups}, Commun.Math. Phys. \textbf{26}, 222--236 (1972).}
	
	\bibitem[Brif and Mann(1996)]{Brif:1996oj}
	C.~Brif and A.~Mann.  \href{https://doi.org/10.1103/PhysRevA.54.4505}
	{\emph{Nonclassical interferometry with intelligent light},  Phys. Rev. A  \textbf{54}, 4505--4518 (1996).}
	
	\bibitem[Vilenkin and Klimyk(1991)]{Vilenkin:1991aa}
	N.~Ja. Vilenkin and A.~U. Klimyk.
	\newblock \emph{Reperesentation of Lie Groups and Special Functions}, volume~1.
	\newblock Springer, Berlin, 1991.
	
	\bibitem[Mehler(1881)]{Mehler:1881aa}
	F.G Mehler.
	\newblock Ueber eine mit den kugel-und Cylinderfunctionen verwandte Function
	und ihre Anwendung in der Theorie der Electricit{\"a}tsvertheilung,
	\newblock \emph{Math. Ann.} \textbf{18}, 161--194 (1881).
	
	\bibitem[Fok(1943)]{Fock:1943aa}
	V.~A. Fock.
	\newblock On the representation of an arbitrary function by an integral
	involving Legendre functions with complex index,
	\newblock \emph{Dokl. Akad. Nauk SSSR}, \textbf{39},  253--256 (1943).
	
	\bibitem[Klimov et~al.(2021)Klimov, Seyfarth, de~Guise, and
	S{\'a}nchez-Soto]{Klimov:2021aa}
	A.~B. Klimov, U.~Seyfarth, H.~de~Guise, and L.~L S{\'a}nchez-Soto. \href{https://doi.org/10.1088/1751-8121/abd7b4}
	{\emph{SU(1, 1) covariant $s$-parametrized maps},  J. Phys. A: Math. Theor.  \textbf{54},  065301 (2021).}

	
	\bibitem[Barut and Girardello(1971)]{Barut:1971yq}
	A.~O. Barut and L.~Girardello. \href{https://doi.org/10.1007/BF01646483}
	{\emph{New ``coherent'' states associated with non-compact groups},  Comm. Math. Phys. \textbf{21},41--55 (1971).}
	
	\bibitem[Baltazar et~al.(2025)Baltazar, Valtierra, and Klimov]{Baltazar:2025aa}
	M.~Baltazar, I.~F. Valtierra, and A.~B. Klimov. \href{https://doi.org/10.1016/j.aop.2025.170208}
	{\emph{Quantum systems in the hyperbolic phase-space: Explicit maps, differential form of the star product and their applications}, 
	Ann. Phys. \textbf{482}, 170208 (2025).}

	
	\bibitem[Bluhm et~al.(1995)Bluhm, Kosteleck{\'y}, and Tudose]{Bluhm:1995}
	R.~Bluhm, V.~A. Kosteleck{\'y}, and B.~Tudose. \href{https://doi.org/10.1103/PhysRevA.52.2234}
	{\emph{Elliptical squeezed states and Rydberg wave packets},  Phys. Rev. A  \textbf{52},2234--2244 (1995).}
	
	\bibitem[{\relax DLMF}()]{NIST:specfun}
	{\relax DLMF}. \href{https://dlmf.nist.gov/}
	{\emph{NIST Digital Library of Mathematical Functions}.}
	
	\bibitem[Widder(1941)]{Widder:1941aa}
	D.~V. Widder.
	\newblock \emph{The Laplace Transform}.
	\newblock Princeton University Press, Princeton, 1941.

	
\end{thebibliography}

\end{document}